\documentstyle[aps,pre,multicol]{revtex}
\begin{document}
\draft
\title{Non-Markovian Persistence and Nonequilibrium Critical Dynamics}
\author{Klaus Oerding$^{1,2}$, Stephen J.~Cornell$^3$, and Alan J.~Bray$^3$} 
\address{
$^1$ Department of Physics, University of Oxford OX1 3NP, UK. \\
$^2$ Institute for Theoretical Physics, University of D\"usseldorf, 
D-40225 D\"usseldorf, Germany. \\
$^3$ Department of Physics and Astronomy, The University, Manchester M13 9PL, 
UK. 
}
\date{February 21, 1997}
\maketitle

\begin{abstract}

The persistence exponent $\theta$ for the global order parameter, $M(t)$,  
of a system quenched from the disordered phase to its critical point 
describes the probability, $p(t) \sim t^{-\theta}$, that $M(t)$ does not 
change sign in the time interval $t$ following the quench. We calculate 
$\theta$ to $O(\epsilon^2)$ for model A of critical dynamics (and to order 
$\epsilon$ for model C) and show that at this order $M(t)$ is a non-Markov 
process. Consequently, $\theta$ is a new exponent. The calculation is 
performed by expanding around a Markov process, using a simplified version 
of the perturbation theory recently introduced by Majumdar and 
Sire [Phys.\ Rev.\ Lett.\ {\bf 77}, 1420 (1996)].   

\end{abstract}
\pacs{}

\begin{multicols}{2}
The `persistence exponent', $\theta$, which characterizes the decay of the 
probability that a stochastic variable exceeds a threshold value 
(typically its mean value) throughout a time interval, has attracted a 
great deal of recent interest 
\cite{DBG,BDG,Stauffer,BKR,DOS,DHP,Boston,MS,MSB,DHZ,MBC}. One of the most 
surprising properties of this exponent is 
that its value is highly non-trivial even in simple systems. For example, 
$\theta$ is irrational for the $q>2$ Potts model in one dimension \cite{DHP} 
(where the fraction of spins that have not changed their state in the time $t$ 
after a quench to $T=0$ decays as $t^{-\theta}$) and is apparently not a 
simple fraction for the diffusion equation \cite{MSB,DHZ} (where the 
fraction of space where the diffusion field has always exceeded its 
mean decays as $t^{-\theta}$).

A recent study of non-equilibrium model A critical dynamics, where a system 
coarsens at its critical point starting from a disordered initial condition, 
looked at the probability $P(t_1, t_2)$ that the global
magnetization does not change sign during the interval $t_1<t<t_2$ 
\cite{MBC}. The persistence exponent for this system is defined 
by $P(t_1, t_2)\sim (t_1/t_2)^\theta$ in the limit $t_2/t_1\to\infty$. 
Explicit results were obtained for the 1D Ising model, the $n \to \infty$ 
limit of the $O(n)$ model, and to order $\epsilon=4-d$ near dimension $d=4$. 
For these systems it was found that the value of $\theta$ was related to the 
dynamic critical exponent $z$, the static critical exponent $\eta$, and 
`nonequilibrium'  exponent $\lambda$ (which describes the decay of the 
autocorrelation with the initial condition, 
$\langle \phi({\bf x},t)\phi({\bf x},0) \rangle \sim t^{-\lambda/z}$) 
by the scaling relation $\theta z = \lambda - d + 1 -\eta/2$. 
This relation may be derived from the assumption that the dynamics is 
Markovian, which is indeed the case for all of the cases considered in 
that paper.

 From a consideration of the structure of the diagrams which appear at 
order $\epsilon^2$ (and higher order), however, it was argued that 
the dynamics of the global order parameter should not be Markovian 
to all orders, implying that the exponent $\theta$ does not obey exactly 
that `Markovian scaling relation' \cite{MBC}. This means that $\theta$ is 
a new exponent. Monte-Carlo simulations in 2 dimensions indeed suggest 
weak violation of the Markov scaling relation \cite{MBC}.

In this Rapid Communication we present an explicit calculation of the
non-Markovian properties of the global order parameter. The nonequilibrium 
magnetization-magnetization correlation function is calculated to
order $\epsilon^2$, and this is then used to calculate $\theta$ to the same 
order, using a perturbative method proposed by Majumdar and Sire (MS) 
\cite{MS}, valid in the vicinity of a Markov process.  The Markov scaling 
relation is shown explicitly to be violated at order $\epsilon^2$, 
so $\theta$ is indeed a new independent exponent.

Before discussing the calculation of $\theta$, however, we provide first 
a simpler, and more transparent, formulation of the perturbation 
theory than that given in MS. In particular the final result, 
Eq.\ (\ref{thetafinal}), does not appear explicitly in MS \cite{Note}.

Let $y(t)$ be a Gaussian stochastic process with zero mean, whose
correlation function obeys dynamical scaling, i.e.\ $\langle
y(t_1)y(t_2)\rangle = t_1^{\alpha}\Phi\left(t_1/t_2\right)$.  Let $T=\ln
t$ and $x(T)=y(t)/\left\langle y^2(t)\right\rangle^{1/2}$.
Then $x(t)$ is a Gaussian {\sl stationary\/} process with zero mean,
i.e.\ its correlation function is translationally invariant, $\langle
x(T_1)x(T_2)\rangle=A(T_2-T_1)$.  Notice that $A(0)=1$ by construction, 
which convention we shall adopt throughout this Communication (in contrast
to that of ref.\ \cite{MS}).  If the persistence probability of $y$
decays algebraically in $t$, then the persistence probability of
$x(T)$ decays as $\sim \exp(-\theta T)$ for $T \to \infty$.

The persistence probability may be expressed as the ratio of two path
integrals, as follows \cite{MS}:
\begin{equation}
  P(x(T')>0; 0<T'<T)=\frac{\int_{x>0}Dx(T)\exp(-S)}
  {\int Dx(T)\exp(-S)},\label{pathint}
\end{equation}
where 
\begin{equation}
  S={1\over 2}\int_0^TdT_1\int_0^TdT_2\, x(T_1)G(T_1, T_2)x(T_2).
\end{equation}
Here $G(T_1,T_2)$ is the matrix inverse of the correlation  
matrix $\langle x(T_1)x(T_2)\rangle\equiv A(T_2-T_1)$.
Notice that $G$ is not simply a function of $T_2-T_1$ (unless we
impose periodic boundary conditions).

In MS this path-integral formalism was used to map the Markov process 
onto a quantum harmonic oscillator in imaginary time, developing the 
perturbation theory in the formalism of quantum mechanics.  We shall 
merely use path integrals as a convenient notation, performing all our 
calculations within the natural framework of stochastic processes.

Let $x^0(T)$ be a stationary Gaussian Markov process, i.e.\ one defined by
\begin{equation}
  \label{lang-ou}
  {dx^0\over dT}=-\mu x^0+\xi(T),
\end{equation}
where $\xi$ is a Gaussian white noise, with $\langle
\xi(T)\xi(T')\rangle = 2\mu\delta(T-T')$.  The noise strength has
been chosen so that the autocorrelation function is 
$A^0(T)=\exp(-\mu T)$.

Suppose the process $x(T)$ is perturbatively close to a Markov
process, in the sense that $G=G^0+\epsilon g$.  Then we can
expand the exponentials in the path integrals in (\ref{pathint}) and 
re-exponentiate, so that to $O(\epsilon)$ the numerator becomes 
\begin{equation}
  \label{gav1}
  \int_{\cal C} Dx(T)e^{-S}
=\int_{\cal C} Dx(T)\exp\left(-S^0-{\epsilon\over 2}\gamma(T)
+ O(\epsilon^2) \right),
\end{equation}
where the subscript ${}_{\cal C}$ represents the constraint $x(T^{\prime}) 
>0$
($0 < T^{\prime} < T$) on the paths in the integral in the numerator of
(\ref{pathint}), and 
\begin{equation}
 \gamma(T)= \int_0^TdT_1\int_0^TdT_2\,
g(T_1,T_2)A_{\cal C}^0(T_1,T_2),
\end{equation}
where
\begin{equation}
  \label{constav}
  A_{\cal C}^0(T_1,T_2)\equiv\frac{\int_{\cal C} Dx(T)\,
 x(T_1)x(T_2)e^{-S^0}}{\int_{\cal C} Dx(T)\,e^{-S^0}}
\end{equation}
is the correlation function for the Markov process, averaged (and
normalized) only over the paths consistent with the constraint ${\cal C}$.
The denominator in (\ref{pathint}) is given by an identical expression,  
except that $A^0_{\cal C}$ is replaced by $A^0$, the unconstrained 
correlation function.

By virtue of the constraint, $A_{\cal C}^0$ will not be 
strictly translationally invariant for finite $T$. In the limit 
$T \to \infty$, however, the double time-integral in (\ref{gav1}) 
reduces to $T$ times an infinite integral over the relative time $T_2-T_1$, 
with $A_{\cal C}^0(T_1, T_2)$ replaced by its stationary limit 
$A_{\cal C}^0(T_2-T_1)$. Similarly, $g$ will be translationally invariant 
in this regime, giving  
\begin{equation}
\gamma(T)\to T\int_{-\infty}^{\infty} (d\omega/2\pi)\,\tilde g(\omega) 
\tilde A_{\cal C}^0(\omega)\ ,\label{dross3}
\end{equation}
where we have used the translational invariance to write the final result in 
Fourier space \cite{Note1}.  Note that the zeroth-order result 
$\int_{x>0} Dx(T)\exp(-S^0)/\int Dx(T)\exp(-S^0)$ is just the persistence 
probability of the stationary Gaussian Markov process $x^0(T)$, which 
decays as $\exp(-\mu T)$ as $T\to\infty$.

Using  (\ref{pathint}), (\ref{gav1}) and (\ref{dross3}), 
we find that the persistence exponent may be written in the form 
\begin{eqnarray}
  \theta&\equiv&\lim_{T\to\infty}-{1\over T}\log\left[P(x(T')>0; 
0<T'<T)\right] \nonumber \\
  &=&\mu + \epsilon \int_0^\infty \frac{d\omega}{2\pi}
  \tilde{g}(\omega)\left[\tilde{A}_{\cal C}^0(\omega)
   -\tilde A^0(\omega)\right] +O(\epsilon^2)\ . 
\label{thetapert}
\end{eqnarray}
where the term in $\tilde{A}^0(\omega)$ is the $O(\epsilon)$ contribution 
from the denominator in (\ref{pathint}), and we have exploited the 
$\omega \to -\omega$ symmetry of the integrand.

We now calculate $A_{\cal C}^0(T)$.  The conditional probability 
$Q(x,T|x_0,0)$ for the stationary Markov process may be obtained 
directly from (\ref{lang-ou}):
\begin{eqnarray}
\lefteqn {Q(x, T|x_0, 0)=}\nonumber\\
&&\hbox to 1 truecm{\hfill}{\left[{1\over 2\pi\left(1-e^{-2\mu T}\right)}
\right]^{1/2}
 \exp\left[-{\left(x-x_0e^{-\mu
   T}\right)^2\over2\left( 1-e^{-2\mu T}\right)}\right]}. 
   \label{greens-ou}
\end{eqnarray}
The conditional probability $Q^+(x_2, T_2|x_1, T_1)$ that the
process goes to $(x_2, T_2)$, given that it started from  $(x_1,
T_1)$, without $x$ ever being negative is given by the
method of images: 
\begin{equation}
  Q^+(2|1)=Q(x_2, T_2|x_1, T_1)-Q(x_2,
  T_2|{ -x_1} , T_1),\label{hardwall}
\end{equation}
where we have adopted an obvious shorthand notation for the
arguments of $Q^+$.

To calculate the joint probability $P^+(x_1, T_1; x_2, T_2)$ that
the process passes through $x_1$ at $T_1$ and $x_2$ at $T_2$, averaged
only over paths where $x(T)$ is always positive, we consider a path
starting at $(x_i, T_i)$ and finishing at $(x_f, T_f)$, passing
through $(x_1, T_1)$ and $(x_2, T_2)$ without ever crossing the
origin.  Then the required stationary limit is 
\begin{equation}
  P^+(x_1, T_1; x_2, T_2)=\lim_{T_i\to-\infty,
    T_f\to\infty}{Q^+(f; 2; 1|i)\over
    Q^+(f|i)}.\label{joint}
\end{equation}
The Markov property means that we can decompose
$Q^+(f; 2; 1|i)=Q^+(f |2)Q^+(2 |1)Q^+(1|i)$.
Using (\ref{greens-ou}) and (\ref{hardwall}) in (\ref{joint}), we find
\begin{eqnarray}
  P^+(x_1, 0; x_2, T)={2\over
    \pi}\left(1-e^{-2\mu T}\right)^{-1/2}x_1x_2\,e^{\mu
    T}\times\phantom{\times}&&\\
\exp\left[-{\left(x_1^2+x_2^2\right)\over 2\left( 1-e^{-2\mu
      T}\right)}\right] \sinh \left({x_1x_2\over 2\sinh \mu
    T}\right). && \label{joint2}
\end{eqnarray}
It is now  straightforward to evaluate the autocorrelation function:
\begin{eqnarray}
  A_{\cal C}^0(T)&=&\int_0^\infty dx_1\int_0^\infty dx_2\,x_1x_2P^+(x_1,
  0; x_2, T)\\ 
  &=&{2\over\pi}\left[3\left(1-e^{-2\mu
    T}\right)^{1/2}\right.\nonumber\\
&&\left. \phantom{{2\over\pi}\left[3\right.}+\left(e^{\mu T}+2e^{-\mu
    T}\right)\sin^{-1}e^{-\mu T}\right].\label{corfunconst}
\end{eqnarray}

Eq.\ (\ref{thetapert}) for $\theta$ can now be expressed as a real-time 
integral as follows. 
We first write $A(T)=A^0(T) + \epsilon a(T)$, and we note that
in Fourier space $\left[\tilde A(\omega)\right]^{-1} = \tilde G(\omega) 
= \tilde G^0(\omega) + \epsilon \tilde g(\omega) $. 
Using $A^0=\exp(-\mu T)$ gives 
$\tilde g(\omega) = - \tilde a(\omega) (\omega^2+\mu^2)^2/4\mu^2$.
Inserting this in (\ref{thetapert}), and transforming to real time, gives
\begin{eqnarray}
\theta & = & \mu - \frac{\epsilon}{4\mu^2} \int_0^\infty dT\,a(T)\,
\left(\mu^2 -\partial_T^2\right)^2 \left[A_{\cal C}^0(T) 
- A^0(T)\right] \nonumber \\
& = & \mu\left\{1-\epsilon
{2\mu\over\pi}\int_0^\infty {a(T)\over
\left[1- \exp (-2\mu
T)\right]^{3/2}} dT\right\}.
  \label{thetafinal}
\end{eqnarray}
The final result is remarkably compact. Since $\epsilon a(T)$ is just 
the perturbation to the Markov correlator $A^0(T) = e^{-\mu T}$, the 
normalization $A(T)=1$ forces $a(0)=0$. This is sufficient to converge 
the integral in (\ref{thetafinal}) provided $a(T)$ vanishes more rapidly 
than $T^{1/2}$. Eq.\ (\ref{thetafinal}) has recently been used to   
calculate persistence exponents for interface growth in a class of 
generalized Edwards-Wilkinson models \cite{krug}.

As was remarked earlier, the problem of non-equilibrium critical
dynamics is Markovian to first order in $\epsilon=4-d$. In the thermodynamic 
limit the global order parameter is Gaussian because, at time $t$, it
is the sum of $[L/\xi(t)]^d$ (essentially) statistically independent 
contributions,  where $L$ is the system size and $\xi \sim t^{1/z}$ is the 
length scale over which critical correlations have been established.
Corrections to the Gaussian distribution can be expressed in terms of higher 
cumulants of the normalized total magnetization $M(t)/\langle M^{2}(t)
\rangle^{1/2}$. Using the translational invariance of the system with respect
to space it is easy to show that for large $L$ the $2N$-point cumulant is 
smaller
by a factor $(t^{1/z}/L)^{(N-1) d}$ compared to the Gaussian part of the
$2N$-point correlation function. The perturbative approach
discussed in the first part of this Communication can therefore be applied.
To calculate the lowest non-Markovian term in $\theta$, we need to
calculate the autocorrelation function of the total magnetization $M(t)$ 
to order $\epsilon^2$, i.e.\ we need to 
calculate the autocorrelation function 
$A(t_1,t_2)= \langle M(t_1)M(t_2) \rangle/
\langle M^2(t_1)\rangle^{1/2} \langle M^2(t_2)\rangle^{1/2}$, 
which in the scaling regime depends only on the ratio $t_2/t_1$. 
The necessary techniques of dynamical field-theory, incorporating the extra 
renormalization associated with the random initial condition (and responsible 
for the nonequilibrium exponent $\lambda$), have been developed by 
Janssen et al \cite{Janssen,J2}. We first consider `model A' dynamics \cite{HH} 
for a nonconserved $O(n)$ model (where  `persistence' is associated with a 
given component of the order parameter for $n>1$).  
The calculation is straightforward in principle (but algebraically 
tedious), and the final result is (with $T = \ln (t_2/t_1)$)
\begin{equation}
  \label{C}
  A(T)=e^{-\mu T}\left[ 1- {3(n+2)\over4(n+8)^2}\epsilon^2
  F_A\left(e^{T}\right) + O(\epsilon^3) \right],
\end{equation}
where $\mu = (\lambda - d + 1 - \eta/2)/z$ from the one-loop calculation 
\cite{MBC} (equivalent to the `Markov scaling relation'), and 
\def\Li{{\rm Li}_2}
\begin{eqnarray*}
 && F_A(x)=\\
&&-\ln{4\over
    3}\left[2\ln(2x)+(x-1)\ln(x-1)-(x+1)\ln(x+1)\right]\\
 &&  -2 \left(\ln 2\right)^2-{\pi^2\over 6} +4\ln 2 -(x-1)\ln\left({x-1\over
    2x}\right)\\
&& +(x+1)\ln\left({x+1\over
    2x}\right) +(x-1)\ln\left({x-1\over 2x}\right)\ln\left({3x-1\over
    2x}\right) \\
&&-(x+1)\ln\left({x+1\over 2x}\right)\ln\left({3x+1\over
    2x}\right)\\
&&-(3x+1)\ln\left({3x+1\over
    2x}\right)+(3x-1)\ln\left({3x-1\over 2x}\right)\\
  &&-{(x-1)\over
    2}\left[\ln\left({3x-1\over 2x}\right)\right]^2
   +{(x+1)\over
    2}\left[\ln\left({3x+1\over
   2x}\right)\right]^2\\
 && -(x-1)\Li\left({x-1\over
   2x}\right)+(x+1)\Li\left({x+1\over
   2x}\right)\\
&& -2(x+1)\Li\left({x+1\over
   4x}\right) +2(x-1)\Li\left({x-1\over 4x
   }\right)\\
&&+(x+1)\Li\left({2x\over 3x+1}\right) 
 -(x-1)\Li\left({2x\over 3x-1}\right),
\end{eqnarray*}
and $\Li(x)\equiv -\int_0^x dt\, \ln(1-t)/t$ is the dilogarithm function.
The function $F_A(e^T)$ is a bounded, monotonically increasing, function 
of $T$ in $(0,\infty)$. It vanishes as $T\ln T$ for $T \to 0$ [satisfying 
the requirement for convergence at $T=0$ of the integral in 
(\ref{thetafinal})], while $F(\infty)=0.057622\ldots$ 

The non-Markov nature of the process $M(t)$ at order $\epsilon^2$ follow 
from the fact that, at this order, $A(T)$ is no longer a simple exponential. 
Substituting $a(T)=A(T)-e^{-\mu T}$ from (\ref{C})
into Eqn.\ (\ref{thetafinal}), using $\mu=(1/2)+O(\epsilon)$, we find
(after some algebra)
\begin{equation}
\theta=\mu\left\{ 1+ {3(n+2)\over 4(n+8)^2}\epsilon ^2 \alpha\right\},
\label{nonmarkov}
\end{equation}
where
\begin{eqnarray*}
\alpha &=& 4\left(\sqrt{2} - 2\sqrt{3} + \sqrt{6}\right) + 8 \sqrt{2} \ln 2
        - 4 \left(\sqrt{2}-1\right) \ln 3 \\
&& - 2 \left(1+2\sqrt{2}\right) \ln\left(3+2\sqrt{2}\right)
       - 14 \ln\left(5+2\sqrt{6}\right) \\
&& + 10 \ln\left(7+4\sqrt{3}\right)
        + 8 \sqrt{2} \ln\left( {4+\sqrt{2}-\sqrt{6}\over 4-\sqrt{2}-\sqrt{6}}
        \right)\\
&&
        - 4 \sqrt{2} \ln\left(
        {2\sqrt{3}-2+\sqrt{2}\over 2\sqrt{3}-2-\sqrt{2}} \right)\\
&=&  0.271577604975\dots
\end{eqnarray*}

This result can be compared with recent simulation data for the Ising 
model in 2 \cite{MBC,Stauffer96,SZ} and 3 \cite{Stauffer96} dimensions. 
For $d=2$, using $\lambda = 1.585 \pm 0.006$ \cite{Grassberger}, and 
$\eta=1/4$ (exact) gives $\mu z = 0.460 \pm 0.006$. Ignoring non-Markov 
corrections, one would obtain $\theta z = \mu z$, smaller than 
the measured value $\theta z = 0.505 \pm 0.020$ (the finite-size scaling 
method used in \cite{MBC} naturally determines the combination $\theta z$
\cite{SZ}).  
The non-Markov correction factor in (\ref{nonmarkov}) is, for $n=1$, 
$(1+0.0075438\ldots\epsilon^2) \simeq 1.030$ for $\epsilon=2$. The 
`improved' estimate for $\theta z$ becomes $0.474 \pm 0.006$, closer to, 
but still somewhat smaller than, the numerical estimate. 

For $d=3$, one has $z=2.032 \pm 0.004$, $\lambda=2.789 \pm 0.006$ 
\cite{Grassberger} and $\eta = 0.032 \pm 0.003$, giving 
$\mu = 0.380 \pm 0.003$. Multiplying by the non-Markov correction factor 
for $\epsilon=1$, i.e.\ $1.0075$, gives $\theta = 0.383 \pm 0.003$, compared 
to the numerical result $\theta \simeq 0.41$ \cite{Stauffer96}. 
A direct expansion to order $\epsilon^2$, using the known expansions 
for $z$, $\lambda$, and $\eta$, gives (specializing to $n=1$) 
$\theta = 1/2 - \epsilon/12 + (\alpha - 2\ln 3)\epsilon^2/72 - 
2\epsilon^2/81 + O(\epsilon^3)$, i.e.\ $\theta \simeq 0.365$ for $d=3$, 
slightly lower than that obtained using the best numerical estimates of 
$z$, $\lambda$ and $\eta$ and only using the $\epsilon$-expansion for 
the non-Markov correction.

A similar approach can be applied to `model C' critical dynamics \cite{HH}, 
in which a nonconserved order parameter field is coupled to a conserved 
density. In this case, one obtains non-Markovian corrections 
already at order $\epsilon$. The autocorrelation function is given by 
(for $n=1$)
\begin{eqnarray}
\label{A_C}
A(T) & = & \exp(-\mu T)\left[1 - \frac{\epsilon}{6}\,F_C(e^T) 
+ O(\epsilon^2)\right] \\
F_C(x) & = & \ln 2 - \frac{x-1}{2x} + x\ln x\nonumber\\
&&-\frac{x-1}{2}\ln(x-1) -\frac{x+1}{2}\ln(x+1). 
\label{F_C}
\end{eqnarray}
Again, $F_C(e^T)$ vanishes like $T\ln T$ for $T \to 0$, while 
$F_C(\infty) = \ln 2 -1/2$. Inserting $a(T) = A(T) - \exp(-\mu T)$ 
from (\ref{A_C}) into (\ref{thetafinal}) gives 
\begin{equation}
\theta = \mu\left[1 + \frac{7-4\sqrt{2}}{12}\epsilon + O(\epsilon^2)\right],
\end{equation}
where $\mu = (\lambda - d + 1 -\eta/2)/z$ as before, but now the 
dynamical exponents $z$ and $\lambda$ take their model-C values 
\cite{HH,Oerding}.

In summary, we have computed to order $\epsilon^2$ the persistence exponent 
$\theta$ for the global order parameter $M(t)$ of models A and C. 
At this order, the dynamics of $M(t)$ are non-Markovian, and $\theta$ is 
a new exponent, not related to the usual static and dynamic exponents. 
The calculation was performed by expanding around a Markov process, 
using a simplified form of the perturbation theory introduced by 
Majumdar and Sire. 

The work of SC and AB was supported by EPSRC grants GR/K53208 and 
GR/J24782 and the work of KO was supported by grant Oe199/1-1 of
the DFG.

\end{multicols}
                                          

\begin{references}
\bibitem{Note} The result (\ref{thetafinal}) can be deduced after some 
algebra from Eqs.\ (4), (5) and (7) of ref.\ \cite{MS}.
\end{references}
\end{document}